\documentclass[doublespacing]{elsart}

\usepackage{amsmath}
\usepackage{graphicx}

\def\onlinecite{\cite}

\begin{document}

\date{\today}

\begin{frontmatter}

\title
{DNA-based tunable THz oscillator}

\author{
A. V. Malyshev$^{a,b,c}$, V. A. Malyshev$^a$, F.\
Dom\'{\i}nguez-Adame$^b$}

\address{$^a$ Centre for Theoretical Physics and Zernike Institute for Advanced
Materials, University of Groningen, Nijenborgh 4, 9747 AG Groningen,
The Netherlands }

\address{$^b$GISC, Departamento de F\'{\i}sica de Materiales, Universidad
Complutense, E-28040 Madrid, Spain}

\address{$^c$Ioffe Physico-Technical Institute, 26
Politechnicheskaya str., 194021 St.Petersburg, Russia}


\corauth[corauthor]{ Corresponding author: A. V. Malyshev, Centre
for Theoretical Physics and Zernike Institute for Advanced
Materials, University of Groningen, Nijenborgh 4, 9747 AG Groningen,
The Netherlands; tel: +31-50-3634784; fax: +31-50-3634947; email:
a.v.malyshev@rug.nl }

\begin{abstract}

The intrinsic double helix conformation of the DNA strands is known to be
the key ingredient of control of the electric current through the DNA by the
perpendicular (gate) electric field. We show theoretically that Bloch
oscillations in the DNA are also strongly affected by such lateral field;
the oscillation frequency splits into a manifold of several generally
non-commensurate frequencies leading to a complicated pattern of the charge
motion. The frequency of the oscillations falls in the THz domain, providing
for a possibility to design a nano-scale source of THz radiation.

\end{abstract}

\begin{keyword}
Nano-electronic devices, field effect devices, DNA



\end{keyword}

\journal{J. Luminescence}
\end{frontmatter}

\newpage

\section{Introduction}
\label{Sec: Introduction}


The helical symmetry of the DNA is almost always neglected when modeling the
charge transport through the DNA-based devices: a DNA is usually considered
as a flat ladder-like sequence (periodic or stochastic) of base
pairs~\cite{Iguchi97,Yamada04,Yamada05,Klotsa05,Gutierrez06}. Very recently
the intrinsic helix conformation of the DNA strands was put forward as the
key ingredient allowing for control of the electric current through the
molecule by the electric field perpendicular to the DNA
axis~\cite{Malyshev07}. In the presence of such a field the helical
conformation leads to an additional periodic modulation of the base pair
energies, which affects strongly the charge transport through the DNA. On
this basis, prototypes of the single-DNA-based field effect transistor and
the Esaki diode analogue have been proposed~\cite{Malyshev07}.

In this contribution, we aim to further exploit the symmetry of the DNA
molecules and demonstrate, by means of numerical simulations, that (i) - a
single dry periodic DNA molecule (such as the ploy(G)-poly(C)) subjected to
a collinear uniform electric field can exhibit Bloch
oscillations~\cite{Lakhno04} and (ii) - because of the helical conformation,
the oscillations become more complex in the presence of the perpendicular
(gating) electric field. The frequency of these oscillations falls in the
THz domain, the region which is in the focus of an intense research nowadays
(see Ref.~\cite{Gragoman04} for a recent overview). The above-mentioned
property provides therefore for a possibility to design a DNA-based tunable
THz oscillator.

\section{Model}    \label{Sec: Model}

Bloch~\cite{Bloch28} and Zener~\cite{Zener34} argued on the theoretical
grounds that an electron, moving in an ideal periodic potential and
subjected to a uniform electric field $F$, is confined within a finite
region because of the Bragg reflection. Due to the confinement, it undergoes
a periodic motion which is characterized by the angular frequency
$\omega_\mathrm{B}=eFa/\hbar$ and a spatial extension $L_\mathrm{B}=W/(eF)$,
where $-e$ is the electron charge, $F$ is the applied electric field, $a$ is
the lattice constant, and $W$ stands for the band width (see also
Ref.~\cite{Ashcroft76}).

For simplicity, we consider here a single-stranded uniform poly(G) helix for
which we assume the conformation parameters of the double-stranded DNA
(dsDNA) in its B form, in particular, the full-twist period of 10 base
molecules. For a periodic dsDNA, such as the poly(G)-poly(C), the picture of
Bloch oscillations is more complicated as compared to the considered case,
however, the physics is very similar. We use the {\em minimum} tight-binding
model, extending it to include a uniform electric field ${\bf E}$ that is
tilted by the angle $\theta$ with respect to the axis of the helix. The
field has therefore both the collinear component, $F=E\cos{\theta}$, and the
lateral (gating) one, $E\sin{\theta}$. The Hamiltonian of our model in site
representation reads:
\begin{equation}
\label{H}
    H  =  \sum_{n=1}^N
    \big |\varepsilon_{n} |n\rangle\langle n| + J\,\sum_{n=1}^{N-1}
    \big |n+1\rangle\langle n| + h.c. \ .
\end{equation}
Here, $|n \rangle$ is the state vector of the $n$-th base molecule and
the corresponding energy $\varepsilon_{n}$ is given by
\begin{equation}
    \varepsilon_{n} = \varepsilon_{n}^{(0)} - U_{\parallel}n -
    U_{\perp} \,\cos{\left(\frac{2\pi n}{10} + \varphi_0\right)}
    \label{e}
\end{equation}
where $\varepsilon_{n}^{(0)}$ is the site energy of the $n$-th base molecule
at zero field. We assume all molecules to be the same, so we set the
unperturbed energies $\varepsilon_{n}^{(0)}$ to zero from now on. The term
$U_{\parallel}n$ describes the linear potential along the helix axes; the
potential drop across a base molecule in the stacking direction is
$U_{\parallel} = ea E\cos{\theta}$, where $a$ is the nearest-neighbor
distance along the helix axis, $U_{\perp} = eEr\sin{\theta}$ is the
potential drop across the helix in the perpendicular direction, $r \approx
1$ nm being the helix radius. The phase $\varphi_0$ which determines the
azimuth of the strand with respect to the field, is set to zero. The term
$J$ in Eq.~(\ref{H}) describes the transfer interaction between the
nearest-neighbor bases; it is chosen to be positive which implies that the
considered charge is a hole~\cite{Senthilkumar05}.
%

The parallel component of the electric field, $E\cos{\theta}$, yields the
potential ramp along the stacking direction, which sets the frequency of the
Bloch oscillations, as in the traditional case. However, Eq.~(\ref{e}) shows
that the helix in the lateral field acquires the additional periodic
modulation of the potential. This modulation leads to the modification of
the electronic structure of the system: the bare energy band splits into
several different minibands, which is crucial for the charge transport
properties~\cite{Malyshev07}. Each such miniband has its own Bloch
frequency, resulting in a more complex overall picture of Bloch oscillations
as we show below. The amplitude of the periodic modulation is controlled by
the magnitude of the perpendicular component of the electric field,
$E\sin{\theta}$, providing for a mechanism to alter the fundamental
properties of the system.

We further solve the time-dependent Schr\"odinger equation (the Planck
constant $\hbar = 1$)
\begin{equation}
    \label{Schroedinger}
    i{\dot\psi}_{n} = \varepsilon_{n} \psi_{n} + J(\psi_{n+1}
    + \psi_{n-1})
\end{equation}
for an electron wave packet $\psi_n$ being initially a narrow Gaussian
centered at an arbitrary lattice site $n_0$:
\begin{equation}
\psi_{n}(0)= A \exp{\left[-\,\frac{(n-n_0)^2}{2}\right]}\ ,
\label{Gaussian}
\end{equation}
where $A$ is the normalization constant. The solution of
Eq.~(\ref{Schroedinger}) can be expressed in terms of the eigenvalues
$\lambda_{\nu}$ and eigenfunctions $\varphi_{\nu n}$ of the
Hamiltonian~(\ref{H}) as follows
\begin{equation}
    \label{Solution}
    \psi_n(t) = \sum_{\nu=1}^N \sum_{m=1}^N e^{-i\lambda_{\nu}t}
    \varphi_{\nu n} \varphi_{\nu m} \psi_m(0) \ ,
\end{equation}
where the eigenfunctions $\varphi_{\nu n}$ are chosen to be real. The
quantities we use to characterize the dynamics of the electron wave packet are
the mean position of the packet (centroid):
\begin{subequations}
\begin{equation}
    x(t)= \sum_{n=1}^{N}(n-n_0)\,|\psi_{n}(t)|^2\ ,
\label{tools1}
\end{equation}
and its Fourier transform:
\begin{equation}
    f(\omega) = \frac{1}{2\pi} \int_{0}^\infty dt \,
    e^{i \omega t} x(t) \ .
\end{equation}
\end{subequations}

\begin{figure}[ht]
 \begin{center}
         \includegraphics[clip,width=.35\textwidth]{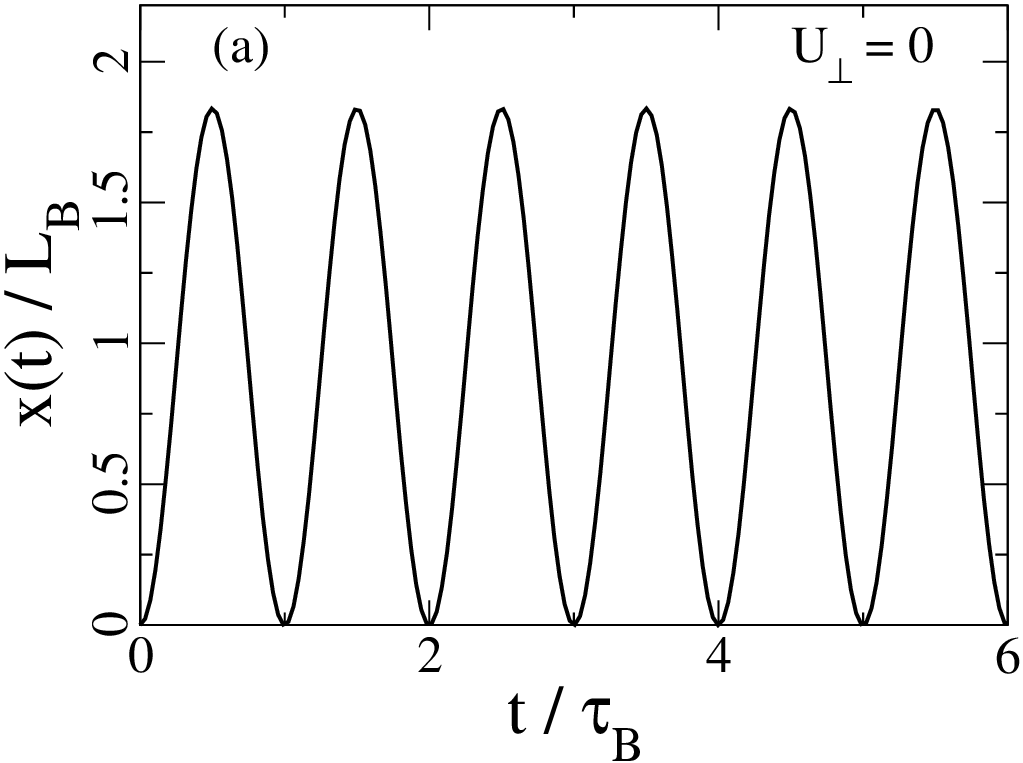}
 \end{center}
 \begin{center}
         \includegraphics[clip,width=.35\textwidth]{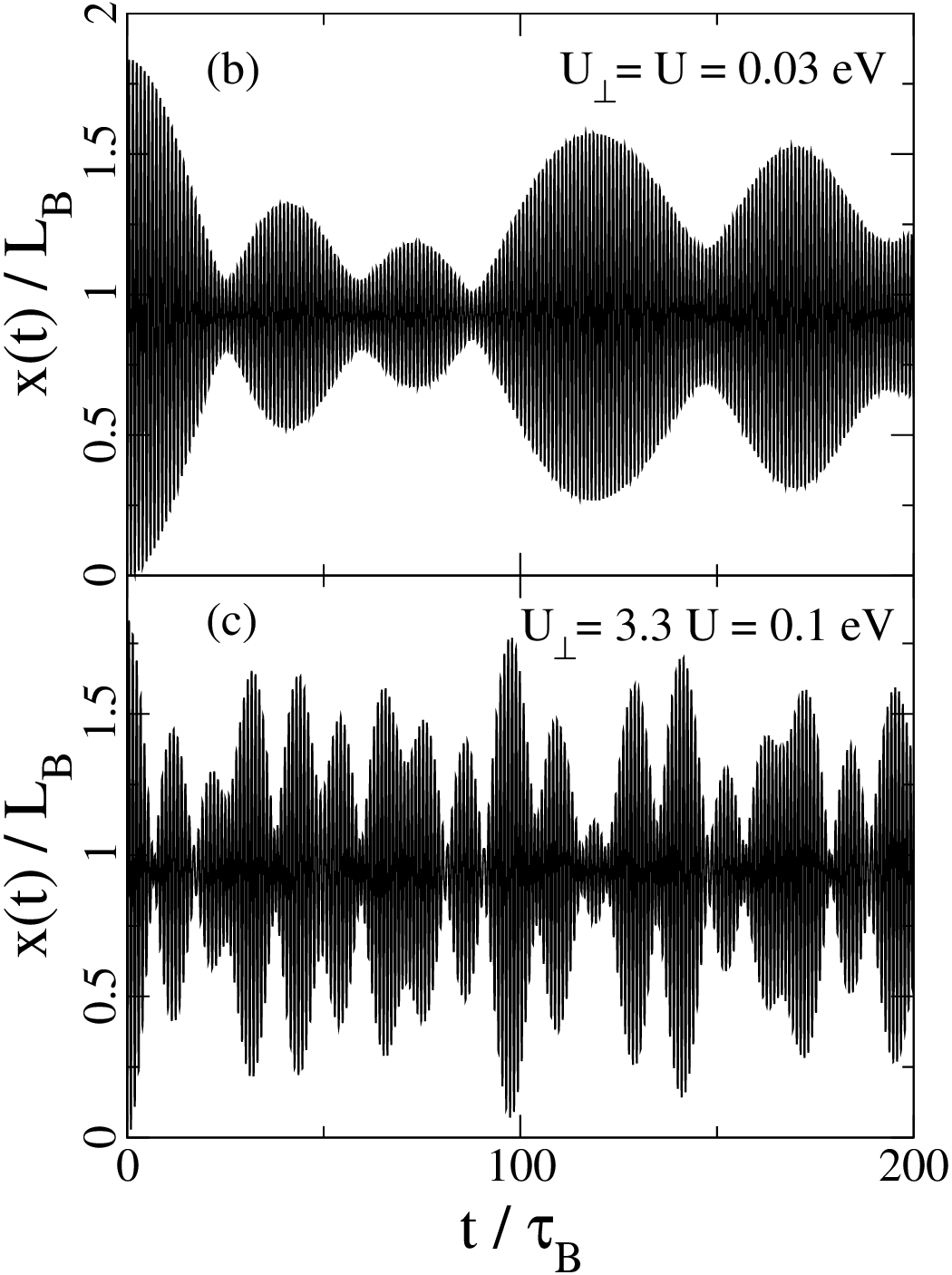}
 \end{center}
    \caption{ Bloch oscillations of the centroid $x(t)$, Eq.~(\ref{tools1}),
        calculated for a single-stranded helix of 101 bases at various
        magnitudes of the gating field. The corresponding gating potentials
        are indicated in the plots.
    }
     \label{Bloch oscillations of the centroid}
\end{figure}

\section{Results and discussions}

In all simulations we used the helix of $N = 101$ bases.
The hopping integral was chosen to be $J = 0.27$ eV (see, e.g.,
Ref.~\onlinecite{Malyshev07}). The initial wave packet was placed in the
middle of the helix, $n_0 = 51$.
%

Figure~\ref{Bloch oscillations of the centroid} displays results of our
calculations of the centroid $x(t)$ for various magnitudes of the gating
potential drop $U_{\perp}$. At zero gating potential (upper panel), the
motion of the centroid represents simple harmonic oscillations with the
period $\tau_B = 2\pi/U_{\parallel}$ and the amplitude $L_B \approx
(4J/U_{\parallel})a$, where $4J$ is the bandwidth, which is in full
correspondence with the standard picture of Bloch oscillations in a linear
chain. The helical symmetry does not affect the oscillations at $U_{\perp} =
0$.

On turning on the gating field, $U_{\perp} \ne 0$,, the motion of the
centroid still manifests an oscillatory behavior, however, a more
complicated one as compared to simple harmonic oscillations (Fig.~\ref{Bloch
oscillations of the centroid} middle and lower panels). The Fourier spectra
$f(\omega)$ of the centroid plotted in Fig.~\ref{Fourier spectra of the
centroid} shed light on the situation. It demonstrates that a nonzero gating
field gives rise to a splitting of the Bloch frequency into a multiplet with
a frequency spacing dependent on the gating potential $U_{\perp}$. The
resulting signals presented in Fig.~\ref{Bloch oscillations of the centroid}
(middle and lower panels) are formed because of the superposition of
harmonic oscillations with several different and generally non-commensurate
frequencies.

\begin{figure}[ht]
 \begin{center}
         \includegraphics[width=.35\textwidth]{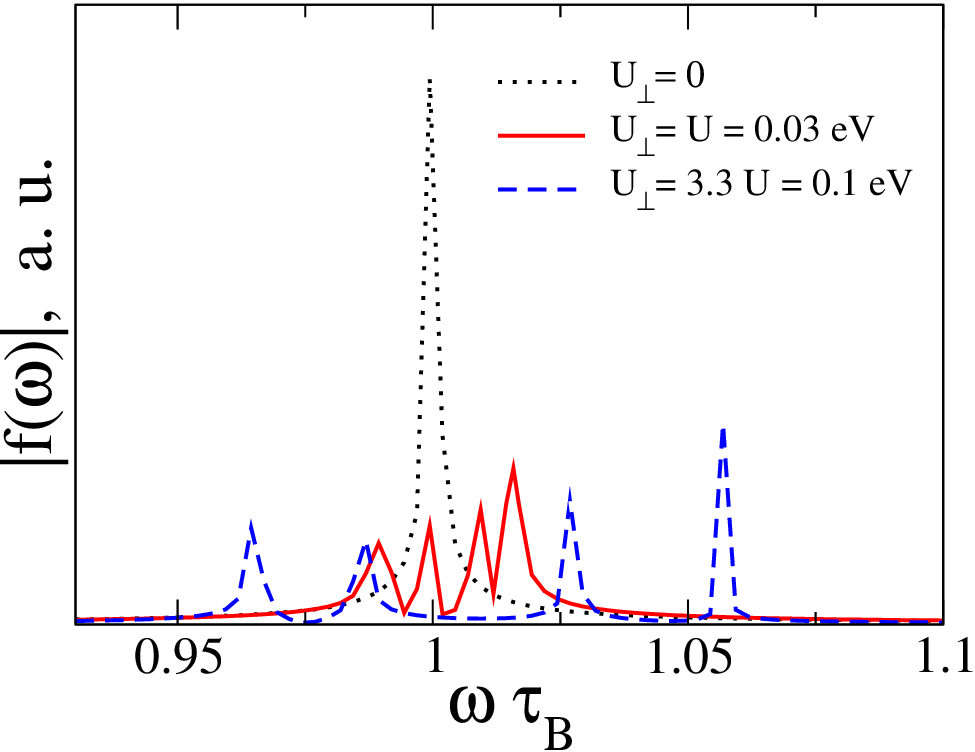}
 \end{center}
    \caption{ Spectra of Bloch oscillations of the centriod depicted in Fig.~\ref{Bloch
        oscillations of the centroid}.
    }
     \label{Fourier spectra of the centroid}
\end{figure}

Finally, for the parameters we use, which are typical for the synthetic dry
DNA, the period of oscillations $\tau_B = 2\pi/U_{\parallel} \sim 1$ ps,
i.e., it falls in the THz domain.

\section{Summary}

In summary, we have demonstrated that the intrinsic helix conformation of
the DNA strands can have strong impact on its radiation properties. The
electric field along the stack direction forces the injected charge to
exhibit Bloch oscillations. In a tilted electric field, however, the
harmonic Bloch oscillations become a superposition of oscillations with
close and generally non-commensurate frequencies which can be tuned by an
external electric field. The frequency of the oscillation falls in the THz
domain. This finding is important for the self-assembled DNA arrays on gold
with the DNA molecules being tilted with respect to the
surface~\cite{Kelley98,Kelley99}. Such arrays may provide a nano-scaled
source of coherent THz radiation.


\begin{thebibliography}{99}












\bibitem{Iguchi97} K. Iguchi, Int. J. Mod. Phys. B {\bf 11}, 2405 (1997);
    {\it ibid.}, {\bf 17}, 2565 (2003); {\it ibid.}, {\bf 18}, 1845 (2004);
    J. Phys. Soc. Jpn. {\bf 70}, 593 (2001).

\bibitem{Yamada04} H. Yamada, Int. J. Mod. Phys. B {\bf 18}, 1697 (2004);
    Phys. Lett. A {\bf 332}, 65 (2004).

\bibitem{Yamada05} H. Yamada, E. B. Starikov, D. Henning, and J. F.
    R. Archilla, Eur. Phys. J. E {\bf 17}, 149 (2005).

\bibitem{Klotsa05} D. Klotsa, R. A. R\"{o}mer, and M. S. Turner, Biophys. J.
    {\bf 89}, 2187 (2005).

\bibitem{Gutierrez06} R. Guti\'errez, S. Mohapatra, D. Kohen, D. Porath,
    and G. Cuniberti, Phys. Rev. B {\bf 74}, 235105 (2006).

\bibitem{Malyshev07} A. V. Malyshev, Phys. Rev. Lett. {\bf 98}, 096801
    (2007).

\bibitem{Gragoman04} D. Dragoman and M. Dragoman, Progr. Quant. Electr.
    {\bf 28}, 1 (2004).

\bibitem{Lakhno04} The possibility of Bloch oscillations in the DNA was
    discussed in V. D. Lakhno and N. S. Fialko, Pis'ma Zh. \'Eksp.  Teor. 
    Fiz. {\bf 79}, 575 (2004)[JETP Lett. {bf 79}, 464 (2004)], however, the
    helical symmetry was not taken into account.

\bibitem{Bloch28} F. Bloch, Z. Phys. \textbf{52}, 555 (1928).

\bibitem{Zener34} C. Zener, Proc. R. Soc. London, Ser. A \textbf{145}, 523
    (1934).

\bibitem{Ashcroft76} N. W. Ashcroft and N. D. Mermin, \emph{Solid State
     Physics\/} (Saunders Colege Publishers, New York, 1976), P.~213.

\bibitem{Senthilkumar05} K. Senthilkumar, F. C. Grozema, C. Fonseca Guerra,
F. M. Bickelhaupt, F. D. Lewis, Yu. A. Berlin, M. A. Ratner, and L. D. A.
Siebbeles, J. Am. Chem. Soc. {\bf 127} 14894 (2005).


\bibitem{Kelley98} S. O. Kelley, J. K. Barton, N. M. Jackson, L. McPherson,
A. Potter, E. M. Spain, M. J. Allen, and M. G. Hill, Langmuir {\bf14},
6781, (1998).

\bibitem{Kelley99} S. O. Kelley, N. M. Jackson, M. G. Hill, and J. K.\
Barton, Angew. Chem. Int. Ed. {\bf 38}, 941 (1999).














\end{thebibliography}
\end{document}